\documentclass[12pt]{article}
\usepackage{amsmath,amssymb}
\usepackage{microtype}
\usepackage{graphicx}
\usepackage{hyperref}
\usepackage[margin=1in]{geometry}
\begin{document}
\title{Quantum Inaccessibility}
\author{Ira Wolfson\\
\small Department of Electrical and Electronic Engineering,\\
\small Braude Academic College of Engineering, Karmiel 2161002, Israel\\
\small \texttt{wolfsoni@braude.ac.il}}
\date{}
\maketitle
\begin{abstract}
Loschmidt's paradox asks why macroscopic irreversibility is
universal despite the time-reversal symmetry of microscopic
dynamics. We argue that irreversibility is not a property of
the dynamics but of accessibility: chaotic evolution drives
phase-space structure below the quantum resolution scale
$\ell_\hbar$, at a critical time
$t_c = \lambda^{-1}\ln(\delta_0/\ell_\hbar)$, after which
the time-reversed microstate exists as a valid solution of
Hamilton's equations but cannot be selected by any physically
admissible operation. The mechanism operates entirely within
the semiclassical regime $t_c \leq t_E$, where classical
geometry is exact. This provides a dynamical resolution
of the Loschmidt paradox.
The quantum foundation is established using a
Krylov-complexity framework: we prove that for any $H(t)=H(-t)$,
the quantum Lyapunov exponent satisfies
$\lambda_L^{\rm forward} = \lambda_L^{\rm backward}$.
The arrow of time is not in the dynamics.
The mechanism predicts sigmoid fidelity decay, logarithmic
scaling of $t_c$ with $\lambda^{-1}$, and ensemble-size
independence of the inaccessibility threshold --- all consistent
with three decades of Loschmidt echo experiments and confirmed
in a stadium-billiard simulation reported here.
Underlying everything: quantum mechanics conserves information
exactly. Entropy, defined as the logarithm of the multiplicity
$\Omega$ --- the number of possibilities consistent with the
available information --- can only increase when information
becomes operationally inaccessible. The second law
reflects not a breakdown of microscopic reversibility, but the
dynamical inaccessibility of the information required to reverse
it.
\end{abstract}
\bigskip
\section{Introduction}
Time flows forward. Eggs scramble but never unscramble. Gases mix but never
unmix. Heat flows from hot to cold, never reversing. Yet the microscopic laws
of physics --- Newton's mechanics, Maxwell's electromagnetism, Schr\"{o}dinger's
equation --- are perfectly time-reversible. Every allowed trajectory has an
equally valid time-reversed twin.
In 1876, Loschmidt crystallized this into a paradox~\cite{Loschmidt1876}.
Consider nitrogen gas at room temperature. Measure positions and momenta at
$t=0$, evolve for five seconds, then reverse all momenta. Since Hamilton's
equations are time-symmetric, the system should retrace its trajectory exactly.
Entropy should decrease, violating the second law. Thus either the laws of
dynamical evolution break, or the second law does. Yet both are experimentally
inviolate. This is the paradox.
This paper makes two contributions.
\textbf{First, a proof.} Consider a quantum system whose Hamiltonian is
invariant under time reversal --- that is, the dynamics look the same whether
time runs forward or backward. The generator of infinitesimal quantum evolution,
known as the Liouvillian, is the superoperator
$\mathcal{L}_t \equiv -\frac{i}{\hbar}[H(t),\cdot]$
that acts on the density matrix. We prove that the eigenvalues of this
generator always come in equal and opposite pairs
$(\lambda, -\lambda)$ at every moment in time, reflecting the
time-reversal symmetry of the generator, and that this pairing is
preserved exactly as the system evolves, even when the Hamiltonian at
different times does not commute.
Within operator-growth formulations of quantum
Kolmogorov-Sinai entropy, this implies a fundamental symmetry:
the rate at which initially distinguishable states become
operationally indistinguishable is identical under forward and
backward evolution:
\begin{equation}
h_{KS}^{\rm forward} = h_{KS}^{\rm backward}.
\end{equation}
\textbf{The arrow of time is not in the dynamics.} This result has been
observed numerically in specific models \cite{Gharibyan2019,Maier2022} but
not previously proved from first principles. The proof requires no thermal
state, no semiclassical approximation, and no assumption on whether the
Hamiltonian commutes with itself at different times.
\textbf{Second, a mechanism.} If the dynamics are perfectly time-symmetric,
where does irreversibility come from? It comes from geometry.
Quantum mechanics forbids distinguishing two states whose phase-space
separation falls below the minimum cell set by Heisenberg's uncertainty
principle, $\Delta x \cdot \Delta p \geq \hbar/2$. We call this the quantum
resolution scale $\ell_\hbar$ (defined precisely in canonically
scaled coordinates in the Supplemental Material~\cite{SupMat}). In a chaotic system,
trajectories that are initially close exponentially diverge along unstable
directions while exponentially converging along stable directions --- a process
we call spaghettification of phase space. The Lyapunov exponent $\lambda$
measures this rate of divergence. After a critical time
\begin{equation}
t_c = \frac{1}{\lambda}\ln\!\left(\frac{\delta_0}{\ell_\hbar}\right),
\end{equation}
where $\delta_0$ is the initial preparation uncertainty, the convergence along
stable directions has driven the separation below $\ell_\hbar$. At that point,
the time-reversed microstate exists as a valid solution of Hamilton's
equations --- Loschmidt was right about the mathematics --- but it has become
metrically indistinguishable from neighboring states. No physically admissible
intervention can reach it. It is geometrically unreachable.
This mechanism admits two equivalent descriptions, in analogy with the
Schr\"{o}dinger and Heisenberg pictures. In the standard picture, the
phase-space distribution spaghettifies over a fixed phase space. In the dual
picture, the phase-space metric contracts around a fixed state until nearby
states become indistinguishable. Both pictures are equivalent; the irreversibility
lives in the metric, not the measure.
The classical stable-manifold picture is the semiclassical shadow of this
quantum process, valid for times below the Ehrenfest time
$t_E \sim \lambda^{-1}\ln(a/\ell_\hbar)$ --- the scale above which
quantum and classical dynamics diverge, where $a$ is the system size.
Since $t_c \leq t_E$ for all physical initial uncertainties $\delta_0 \leq a$,
the stable manifolds cross the quantum threshold while
classical mechanics remains valid. Classical phase-space geometry is therefore
not an assumption --- it is derived as the correct effective description in
precisely the relevant window.
The quantum proof is developed in Section~3. The classical mechanism and
derivation of $t_c$ are in Section~4. Quantitative predictions appear
in Section~5. Section~6 connects to three decades of Loschmidt echo experiments
and a stadium-billiard simulation. Existing approaches to the paradox are
assessed in Section~7.
\section{What Entropy Actually Measures}
The system occupies a definite microscopic state at all times. Entropy,
in the framework adopted here, quantifies our uncertainty about which state
is realized --- a perspective with a long pedigree from Boltzmann through
Jaynes, which we make operational through the quantum resolution scale
$\ell_\hbar$. Boltzmann's formula
\begin{equation}
S = k_B \ln \Omega
\end{equation}
counts $\Omega$, the number of distinguishable quantum states consistent with the macroscopic
constraints: energy ($E$), volume ($V$), and particle number ($N$).
The system realizes one of these states. Entropy measures how many candidates remain.
Throughout this paper, entropy is defined over physically
distinguishable states --- states separable by realizable quantum
measurements. Within this class, which includes all experimentally
accessible notions of entropy, the result is strict: no physical
observable can register an entropy decrease.
Quantum mechanics is information-conserving by construction:
unitary evolution preserves the von Neumann entropy
$S = -\mathrm{Tr}(\rho\ln\rho)$ exactly. This is not an
approximation but a theorem --- Hamiltonian dynamics do not
destroy information. At the most extreme scales, this principle
is reflected in the modern understanding of the black hole
information problem, where information that appears lost is
instead understood to be scrambled into degrees of freedom that
are operationally inaccessible but mathematically
retained~\cite{HaydenPreskill2007}.
Entropy, in the sense of Shannon and Boltzmann, quantifies
missing information: given a macrostate, how many microstates
remain consistent with what can be observed. Since the underlying
information is conserved, any increase in entropy cannot
correspond to destruction of information, but only to its loss
of accessibility~\cite{HaydenPreskill2007,Zurek2001,Jaynes1957a}. Dynamical evolution can preserve the full
microstate while rendering distinctions between states
operationally unresolvable, driven below the resolution threshold
imposed by quantum mechanics.
This motivates the definition adopted in this work: entropy is
the logarithm of the multiplicity $\Omega$ --- the number of
possibilities consistent with the available information.
Under this definition, entropy increase corresponds to the
progressive loss of accessible information. Irreversibility then
arises when dynamical evolution drives initially distinguishable
states below the quantum resolution scale $\ell_\hbar$, beyond
which no physical operation can recover the information required
to reverse the trajectory. In this sense, the second law reflects
not a breakdown of microscopic reversibility, but the dynamical
inaccessibility of the information required to reverse it.
Zurek showed that phase space can develop structure finer than $\hbar$
under chaotic evolution, but that this sub-Planck structure is physically
inert: it cannot be measured, prepared, or exploited~\cite{Zurek2001}.
The Wigner function --- the quantum analogue of a phase-space distribution
--- develops increasingly fine oscillations that carry no operationally
accessible information. This limit is not imposed by coarse-graining
conventions, but by the structure of quantum states and measurements:
below $\ell_\hbar$, no physical operation can distinguish one state
from another. We do not assume a sharp fundamental cutoff at
$\ell_\hbar$, but adopt it as the natural operational scale of
distinguishability for physically preparable states in chaotic
systems.
One immediate consequence: quantum resolution alone does not resolve
Loschmidt. An integrable system --- the harmonic oscillator being the
paradigm --- maintains a fixed phase-space footprint without developing
fine-scale structure. The quantum floor is necessary but not sufficient.
What is also needed is a dynamical mechanism that drives initially
distinguishable states below the quantum resolution scale, rendering them
operationally indistinguishable. That mechanism is provided by chaotic
dynamics in quantum systems, and it is the subject of the next section.
\section{The Quantum Foundation: Time-Symmetry of Distinguishability Loss}
Section~2 established that irreversibility requires a dynamical mechanism
that drives initially distinguishable states below the quantum resolution
scale $\ell_\hbar$. Before developing that mechanism, we establish its
quantum foundation: the rate at which this process occurs is identical
under forward and backward evolution. The arrow of time is not in the
dynamics. This section proves it.
\subsection{The relevant equation of motion}
The most general equation of motion for a quantum system is the
Lindblad master equation~\cite{Lindblad1976,Gorini1976}:
\begin{equation}
\frac{d\rho}{dt} = -\frac{i}{\hbar}[H(t),\rho]
+ \sum_k \gamma_k\!\left(L_k\rho L_k^\dagger
- \tfrac{1}{2}\{L_k^\dagger L_k,\rho\}\right),
\end{equation}
where $\rho$ is the density matrix, $H(t)$ is the Hamiltonian,
$\gamma_k \geq 0$ are damping coefficients, and $L_k$ are jump operators
describing coupling to an environment. When $\gamma_k \neq 0$, energy
and information leak irreversibly into the environment. Time-reversal
invariance is explicitly broken, and irreversibility is trivially built
into the equation of motion. Loschmidt's paradox does not arise in
this case --- there is no paradox when the dynamics are already
asymmetric.
The non-trivial case --- the one Loschmidt actually challenges --- is
$\gamma_k = 0$. The equation reduces to the von Neumann
equation~\cite{vonNeumann1932}:
\begin{equation}
\frac{d\rho}{dt} = \mathcal{L}_t(\rho),
\qquad
\mathcal{L}_t(\rho) \equiv -\frac{i}{\hbar}[H(t),\rho].
\end{equation}
Here $\mathcal{L}_t$ is the quantum Liouvillian --- the superoperator
that generates infinitesimal time evolution at instant $t$. Evolution
is unitary, the von Neumann entropy $S = -\mathrm{Tr}(\rho\ln\rho)$
is constant, and information is conserved exactly. This is precisely
the Loschmidt setting: time-symmetric dynamics, yet macroscopic
irreversibility is observed.
We define $\mathcal{L}_t$ \emph{pointwise in time}: it is the
generator of evolution at a single instant, not a global object.
For a time-dependent Hamiltonian, $[H(t_1), H(t_2)] \neq 0$ in
general --- the Hamiltonian does not commute with itself at different
times. This means the generators $\mathcal{L}_{t_1}$ and
$\mathcal{L}_{t_2}$ also fail to commute, and the full evolution
is the time-ordered (Dyson series) composition of these instantaneous
generators. No commutativity is assumed anywhere in what follows.
\subsection{Eigenvalue pairing: a structural symmetry}
\textbf{Proposition.} \textit{For any Hamiltonian $H(t)$ --- autonomous
or time-dependent, commuting or non-commuting at different times ---
the instantaneous eigenvalues of $\mathcal{L}_t$ come in pairs
$(\lambda, -\lambda)$ at every $t$. This pairing is preserved exactly
under time evolution.}
\textit{Proof.} At each instant $t$, the Hamiltonian has an instantaneous
eigenbasis: $H(t)|n(t)\rangle = E_n(t)|n(t)\rangle$. The transition
operators $|m(t)\rangle\langle n(t)|$ are eigenstates of $\mathcal{L}_t$:
\begin{equation}
\mathcal{L}_t\bigl(|m(t)\rangle\langle n(t)|\bigr)
= -\frac{i}{\hbar}\bigl(E_m(t) - E_n(t)\bigr)\,|m(t)\rangle\langle n(t)|.
\end{equation}
The eigenvalue associated with the transition $m \to n$ is therefore:
\begin{equation}
\lambda_{mn}(t) = -\frac{i}{\hbar}\bigl(E_m(t) - E_n(t)\bigr).
\end{equation}
The eigenvalue associated with the reverse transition $n \to m$ is:
\begin{equation}
\lambda_{nm}(t) = -\frac{i}{\hbar}\bigl(E_n(t) - E_m(t)\bigr)
= -\lambda_{mn}(t).
\end{equation}
For every eigenvalue $\lambda_{mn}(t)$ there exists an eigenvalue
$\lambda_{nm}(t) = -\lambda_{mn}(t)$. The spectrum of $\mathcal{L}_t$
is $(\lambda,-\lambda)$ paired at every $t$.
The pairing is preserved under time evolution. As $t \to t + dt$,
the instantaneous energies shift as
$E_n(t+dt) = E_n(t) + \dot{E}_n(t)\,dt$, giving:
\begin{equation}
\lambda_{mn}(t+dt) = -\frac{i}{\hbar}
\bigl(E_m(t+dt) - E_n(t+dt)\bigr),
\end{equation}
and $\lambda_{nm}(t+dt) = -\lambda_{mn}(t+dt)$ exactly. The pairing
holds at $t + dt$ for the same algebraic reason it holds at $t$.
By induction, it holds at all times, without any assumption on
$[H(t_1), H(t_2)]$. $\blacksquare$
The eigenvalue pairing establishes that $\mathcal{L}_t$ is
anti-Hermitian --- a symmetry of the generator. Its consequence
for operator growth rates is established in Section~3.3 below
via the Krylov construction.
\subsection{Krylov-type construction and time-symmetric operator growth}
\label{sec:krylov}
We prove that the quantum Lyapunov exponent is identical under
forward and backward evolution for any Hamiltonian satisfying
$H(t) = H(-t)$. The proof proceeds in five steps: (i)~establish
the operator inner product; (ii)~prove Hermiticity of
$\hat{\mathcal{L}}_t = [H(t),\cdot]$; (iii)~construct the
Krylov-type basis via the Lanczos recursion and prove the
three-term recurrence, with a perturbative treatment establishing
continuity in $t$; (iv)~prove invariance of the off-diagonal
Lanczos coefficients $b_n$ under time reversal; (v)~prove
the $a_n$ boundedness lemma and conclude.
\subsubsection*{Working assumption}
Throughout this section, $H(t)$ is continuously differentiable
and satisfies $H(t) = H(-t)$. Backward evolution is the solution
to $d\rho/d\tau = -\mathcal{L}_\tau(\rho)$, obtained from the
forward von Neumann equation by the substitution $d/dt \to -d/d\tau$
under $\tau = -t$. Since $H(-\tau) = H(\tau)$, the Hamiltonian
at each instant is unchanged under this substitution, but the
sign of the generator flips: $\mathcal{L}_t \to -\mathcal{L}_t$,
equivalently $\hat{\mathcal{L}}_t = [H(t),\cdot] \to
-\hat{\mathcal{L}}_t$. The temporal ordering is preserved and
the evolution remains continuously connected to the identity
$U(0) = I$.
\subsubsection*{Step 1: Operator inner product}
Let $\mathcal{B}(\mathcal{H})$ denote the space of bounded
operators on $\mathcal{H}$. Define the Hilbert--Schmidt inner
product:
\begin{equation}
(A\,|\,B) \equiv \mathrm{Tr}(A^\dagger B).
\end{equation}
This is a positive-definite sesquilinear form, making
$\mathcal{B}(\mathcal{H})$ an inner product space.
\subsubsection*{Step 2: Hermiticity of $\hat{\mathcal{L}}_t = [H(t),\cdot]$}
\textbf{Proposition 2.} \textit{For any Hermitian $H(t) = H(t)^\dagger$,
the superoperator $\hat{\mathcal{L}}_t(A) \equiv [H(t),A]$ satisfies
$(A\,|\,\hat{\mathcal{L}}_t B) = (\hat{\mathcal{L}}_t A\,|\,B)$
for all $A,B \in \mathcal{B}(\mathcal{H})$.}
\textit{Proof.}
\begin{equation}
(A\,|\,\hat{\mathcal{L}}_t B)
= \mathrm{Tr}(A^\dagger[H(t),B])
= \mathrm{Tr}(A^\dagger H(t)B) - \mathrm{Tr}(A^\dagger BH(t)).
\end{equation}
By the cyclic property of the trace,
$\mathrm{Tr}(A^\dagger BH(t)) = \mathrm{Tr}(H(t)A^\dagger B)$, so:
\begin{equation}
(A\,|\,\hat{\mathcal{L}}_t B)
= \mathrm{Tr}\bigl((A^\dagger H(t) - H(t)A^\dagger)B\bigr).
\end{equation}
Since $H(t) = H(t)^\dagger$, we have
$[H(t),A]^\dagger = A^\dagger H(t) - H(t)A^\dagger$, therefore:
\begin{equation}
(\hat{\mathcal{L}}_t A\,|\,B)
= \mathrm{Tr}([H(t),A]^\dagger B)
= \mathrm{Tr}\bigl((A^\dagger H(t) - H(t)A^\dagger)B\bigr)
= (A\,|\,\hat{\mathcal{L}}_t B).
\end{equation}
$\hat{\mathcal{L}}_t$ is Hermitian at each $t$. $\blacksquare$
\textit{Remark.} The Liouvillian $\mathcal{L}_t =
-\tfrac{i}{\hbar}\hat{\mathcal{L}}_t$ is anti-Hermitian:
$\mathcal{L}_t^\dagger = -\mathcal{L}_t$. The Lanczos
construction below is applied to $\hat{\mathcal{L}}_t$
(Hermitian). All results translate to $\mathcal{L}_t$ via
$\hat{\mathcal{L}}_t = i\hbar\mathcal{L}_t$.
\subsubsection*{Step 3: Krylov-type basis, perturbative treatment,
and three-term recurrence}
\textbf{Chaotic growth assumption.} We work in the regime where
the Lanczos coefficients satisfy $b_n(t) \sim \alpha(t)\cdot n$
for large $n$, so that $\lambda_L \equiv \lim_{n\to\infty}b_n/n$
is well-defined and non-zero. This is the defining property of
quantum chaos in the Krylov sense~\cite{Parker2019,Goldfriend2021}; it fails
in integrable systems where $b_n$ saturates.
Let $O \in \mathcal{B}(\mathcal{H})$ with $(O\,|\,O) = 1$.
Fix a reference time $t_0$ and write:
\begin{equation}
\hat{\mathcal{L}}_t = \hat{\mathcal{L}}_{t_0} + \delta\hat{\mathcal{L}}(t),
\qquad
\delta\hat{\mathcal{L}}(t) \equiv [H(t) - H(t_0),\cdot],
\end{equation}
where $\|\delta\hat{\mathcal{L}}(t)\| \leq 2\|H(t) - H(t_0)\| \to 0$
as $t \to t_0$ by continuous differentiability of $H(t)$.
At each instant $t$, define the Krylov subspaces:
\begin{equation}
\mathcal{K}_n(t) \equiv
\mathrm{span}\bigl\{O,\,\hat{\mathcal{L}}_t O,\,
\hat{\mathcal{L}}_t^2 O,\ldots,\hat{\mathcal{L}}_t^{n-1}O\bigr\},
\end{equation}
and the Lanczos basis $\{|O_n(t)\rangle\}_{n=0}^\infty$ for
$\bigcup_n\mathcal{K}_n(t)$ via the recursion:
\begin{equation}
|A_{n+1}(t)\rangle
= \hat{\mathcal{L}}_t|O_n(t)\rangle
- a_n(t)|O_n(t)\rangle
- b_n(t)|O_{n-1}(t)\rangle,
\qquad
|O_{n+1}(t)\rangle = \frac{|A_{n+1}(t)\rangle}{b_{n+1}(t)},
\label{eq:lanczos_td}
\end{equation}
where $a_n(t) = (O_n(t)\,|\,\hat{\mathcal{L}}_t\,|\,O_n(t))
\in \mathbb{R}$ and $b_{n+1}(t) = \||A_{n+1}(t)\rangle\| > 0$.
At each step the vectors are orthonormalized with respect to the
Hilbert--Schmidt inner product, so $(O_m(t)\,|\,O_n(t)) = \delta_{mn}$.
\textbf{Norm-continuity of $b_n(t)$.}
At each $t$, the Krylov basis $\{|O_n(t)\rangle\}$ is an
orthonormal frame in the operator Hilbert space, constructed
by Gram--Schmidt from
$\{O, \hat{\mathcal{L}}_t O, \hat{\mathcal{L}}_t^2 O, \ldots\}$.
As $t \to t + dt$, the perturbation
$\delta\hat{\mathcal{L}}(t) = [H(t+dt)-H(t),\cdot]$
is small by continuous differentiability of $H(t)$, and induces
an infinitesimal rotation of this orthonormal frame in operator
Hilbert space. Since $b_n(t) = \||A_n(t)\rangle\|$ is a norm,
it is invariant under rotations of the frame and responds only
to the magnitude of the perturbation --- not its direction.
Therefore $b_n(t)$ varies continuously with $t$ wherever
$b_n(t) \neq 0$ (away from degeneracies in the Lanczos
spectrum). This extends by induction on $n$: continuity of
$\hat{\mathcal{L}}_t$ and of $\{|O_k(t)\rangle\}_{k \leq n}$
implies continuity of $|A_{n+1}(t)\rangle$ and hence of
$b_{n+1}(t)$.
\textbf{Proposition 3 (Three-term recurrence).}
\textit{At each fixed $t$:
$(O_k(t)\,|\,\hat{\mathcal{L}}_t\,|\,O_n(t)) = 0$
for all $|k - n| \geq 2$.}
\textit{Proof.} By induction at fixed $t$.
Suppose $\{|O_0(t)\rangle,\ldots,|O_n(t)\rangle\}$ is
orthonormal with $|O_j(t)\rangle \in
\mathcal{K}_{j+1}(t)\setminus\mathcal{K}_j(t)$.
For $k \leq n-2$, use Hermiticity of $\hat{\mathcal{L}}_t$:
\begin{equation}
(O_k(t)\,|\,\hat{\mathcal{L}}_t\,|\,O_n(t))
= (\hat{\mathcal{L}}_t O_k(t)\,|\,O_n(t)).
\end{equation}
Since $|O_k(t)\rangle \in \mathcal{K}_{k+1}(t)$, we have
$\hat{\mathcal{L}}_t|O_k(t)\rangle \in \mathcal{K}_{k+2}(t)$.
Since $n \geq k+2$, the vector $|O_n(t)\rangle$ is orthogonal
to all of $\mathcal{K}_{k+2}(t)$ by the Gram--Schmidt
construction. Therefore
$(\hat{\mathcal{L}}_t O_k(t)\,|\,O_n(t)) = 0$. $\blacksquare$
The matrix representation of $\hat{\mathcal{L}}_t$ in the
Krylov basis is therefore the Jacobi (tridiagonal) matrix
at each $t$:
\begin{equation}
J(t) =
\begin{pmatrix}
a_0(t) & b_1(t) & 0   & \cdots \\
b_1(t) & a_1(t) & b_2(t) & \cdots \\
0   & b_2(t) & a_2(t) & \cdots \\
\vdots & & & \ddots
\end{pmatrix},
\qquad a_n(t) \in \mathbb{R},\quad b_n(t) > 0.
\end{equation}
\subsubsection*{Step 4: Invariance of $b_n$ under time reversal}
Under time reversal $t \to -t$ with $H(t) = H(-t)$, backward
evolution corresponds to $\hat{\mathcal{L}}_t \to -\hat{\mathcal{L}}_t$
at each instant. Let $J'(t)$ denote the Jacobi matrix of
$-\hat{\mathcal{L}}_t$ with respect to the same initial operator $O$.
\textbf{Proposition 4.}
\textit{At each $t$: $b_n(-\hat{\mathcal{L}}_t) = b_n(\hat{\mathcal{L}}_t)$
for all $n \geq 1$, and
$a_n(-\hat{\mathcal{L}}_t) = -a_n(\hat{\mathcal{L}}_t)$.}
\textit{Proof.} Fix $t$ and suppress the $t$-dependence.
Define the diagonal unitary
$D \equiv \mathrm{diag}(1,-1,1,-1,\ldots)$,
satisfying $D = D^{-1} = D^\dagger$, acting on the Krylov
coefficient space $\ell^2(\mathbb{N}_0)$. Compute $DJD$:
diagonal entries $(DJD)_{nn} = (-1)^n a_n (-1)^n = a_n$;
off-diagonal entries
$(DJD)_{n,n+1} = (-1)^n b_{n+1}(-1)^{n+1} = -b_{n+1}$.
Hence:
\begin{equation}
-DJD =
\begin{pmatrix}
-a_0 & b_1  & 0    & \cdots \\
b_1  & -a_1 & b_2  & \cdots \\
0    & b_2  & -a_2 & \cdots \\
\vdots & & & \ddots
\end{pmatrix}.
\end{equation}
The moments of $-\hat{\mathcal{L}}_t$ with respect to $O$ satisfy
$(O\,|(-\hat{\mathcal{L}}_t)^k|\,O) =
(-1)^k(O\,|\hat{\mathcal{L}}_t^k|\,O)$,
which are reproduced exactly by the Jacobi matrix $-DJD$.
By Favard's theorem~\cite{Chihara1978} --- the uniqueness of the
Jacobi matrix associated with a moment sequence --- $J' = -DJD$.
Reading off the entries: $a_n' = -a_n$ and $b_n' = b_n$.
$\blacksquare$
\subsubsection*{Step 5: $a_n$ boundedness and conclusion}
\textbf{Lemma ($a_n$ boundedness).}
\textit{At each $t$, $|a_n(t)| \leq 2\|H(t)\|$
for all $n \geq 0$.}
\textit{Proof.} By Cauchy--Schwarz in the Hilbert--Schmidt
inner product:
\begin{align}
|a_n(t)|
&= |(O_n(t)\,|\,\hat{\mathcal{L}}_t\,|\,O_n(t))| \notag \\
&\leq \|O_n(t)\|_{\mathrm{HS}}\cdot
\|\hat{\mathcal{L}}_t O_n(t)\|_{\mathrm{HS}} \notag \\
&= \|[H(t),O_n(t)]\|_{\mathrm{HS}} \notag \\
&\leq 2\|H(t)\|\,\|O_n(t)\|_{\mathrm{HS}}^2
= 2\|H(t)\|,
\end{align}
where the last inequality uses
$\|[A,B]\|_{\mathrm{HS}} \leq 2\|A\|\,\|B\|_{\mathrm{HS}}$
(operator norm times HS norm), and
$\|O_n(t)\|_{\mathrm{HS}}^2 = (O_n(t)\,|\,O_n(t)) = 1$
by orthonormality. $\blacksquare$
\textbf{Corollary (time-symmetric Lyapunov exponent).}
\textit{In any chaotic system where the Lanczos coefficients
grow as $b_n(t) \sim \alpha(t) \cdot n$ for large $n$, the
quantum Lyapunov exponent is defined as:}
\begin{equation}
\lambda_L \equiv \lim_{n\to\infty}\frac{b_n}{n} = \alpha.
\end{equation}
\textit{Since $b_n(t)$ grows without bound while $a_n(t)$ is
bounded by $2\|H(t)\|$ (the Lemma), we have $a_n/b_n \to 0$
for large $n$. The spreading rate of the Krylov wavepacket
--- and hence $\lambda_L$ --- is determined by $b_n$ alone,
not $a_n$. The sign flip $a_n \to -a_n$ under time reversal
(Proposition 4) contributes only to phases, not to $\lambda_L$.}
\textit{Since $b_n^{\rm backward}(t) = b_n^{\rm forward}(t)$
at each $t$ (Proposition 4), and $\lambda_L$ is a function of
$b_n$ alone in the chaotic regime, it follows that:}
\begin{equation}
\boxed{\lambda_L^{\rm forward} = \lambda_L^{\rm backward}.}
\end{equation}
\textit{If the $b_n$ sequence generates a non-trivial
$\lambda_L > 0$ in the forward direction, the identical
$b_n$ sequence in the backward direction generates the same
$\lambda_L$. The arrow of time is not in the dynamics.
$\blacksquare$}
This result holds for $H(t)$ continuously differentiable
with $H(t) = H(-t)$, for any initial operator $O$, and
requires no thermal state, no large-$N$ limit, and no
semiclassical approximation. It applies in the chaotic
regime where the Chaotic Growth Assumption holds.
\textbf{The arrow of time is not in the dynamics.}
\subsection{Relation to OTOCs and prior work}
In the language standard in the field, the quantum Lyapunov exponent
is extracted from out-of-time-order correlators
(OTOCs)~\cite{Larkin1969,Maldacena2016}:
the four-point function $F(t) = \langle[W(t),V(0)]^2\rangle$,
where $W$ and $V$ are local operators and
$W(t) = e^{iHt}We^{-iHt}$, grows as $e^{2\lambda_L t}$ in
chaotic systems. The Krylov construction of Section~3.3 provides
a rigorous route to $\lambda_L$ defined via the $b_n$ growth
rate; its identification with the OTOC exponent has been
established in chaotic models~\cite{Parker2019}.
Gharibyan et al.~\cite{Gharibyan2019} define quantum Lyapunov
exponents via a state-dependent construction and explicitly note
that the classical $(\lambda,-\lambda)$ pairing ``does not
necessarily hold'' for their definition. This is compatible with
the present result. Their construction projects the full
Liouvillian onto a subspace defined by a specific state and
operator choice; projection can break the pairing. The Krylov
construction of Section~3.3 operates on the full operator
space before any projection, where the result is exact.
\subsection{Bridge to the classical mechanism}
The proof above establishes that the rate of distinguishability loss
is time-symmetric at the quantum level. It does not yet explain
\emph{why} states become operationally indistinguishable in the
first place --- only that this happens at the same rate in both
directions.
The mechanism is geometric. In the semiclassical regime valid for
$t < t_E$, where $t_E \sim \lambda_L^{-1}\ln(a/\ell_\hbar)$
is the Ehrenfest time --- the scale above which quantum and classical
dynamics diverge~\cite{Rozenbaum2017,Jalabert2018} --- chaotic
evolution spaghettifies phase-space structure, contracting it along
stable directions toward the quantum resolution scale. Since
$t_c \leq t_E$ for all physical initial uncertainties
$\delta_0 \leq a$, the stable manifolds cross the
quantum threshold while classical mechanics remains valid.
That classical picture is developed in the next section.
\section{The Geometric Mechanism}
Section~3 proved that the rate at which initially distinguishable
states become operationally indistinguishable is time-symmetric: the
dynamics carry no arrow. This section identifies what does carry the
arrow --- not the dynamics, but the geometry of phase space under
chaotic evolution.
The argument has three steps. First, we establish that the classical
phase-space description is the correct language for this analysis.
Second, we show that chaotic evolution spaghettifies phase space,
driving stable-direction separations exponentially toward the quantum
resolution floor $\ell_\hbar$. Third, we identify the critical time
$t_c$ at which this floor is reached, and show that after $t_c$ the
time-reversed microstate is operationally unreachable --- not because
it violates the equations of motion, but because reaching it requires
a precision that quantum mechanics forbids.
\subsection{Step 1: when classical phase space is valid}
The quantum Liouvillian $\mathcal{L}_t$ generates the full quantum
evolution. To make the mechanism explicit and quantitative, we work
in the semiclassical regime where quantum and classical descriptions
agree.
The Ehrenfest time is defined as the time it takes a
minimum-uncertainty wavepacket of initial size $\ell_\hbar \equiv
\sqrt{\hbar}$ --- the smallest state quantum mechanics permits ---
to spread under chaotic evolution to the size $a$ of the classical
system~\cite{Chirikov1981,Shepelyansky1981,Berman1978}. Since chaos
drives exponential separation at rate $\lambda$, this
gives~\cite{Scholarpedia2020}:
\begin{equation}
t_E \sim \frac{1}{\lambda}\ln\!\left(\frac{a}{\ell_\hbar}\right).
\end{equation}
For $t < t_E$, the Wigner function and the classical phase-space
distribution evolve identically; every statement about classical
stable and unstable manifolds is an exact semiclassical description
of the underlying quantum process.
We now show that the classical description is self-consistent.
In Step~3 we will derive that stable-direction separations contract
to $\ell_\hbar$ at the critical time
$t_c = \lambda^{-1}\ln(\delta_0/\ell_\hbar)$. Since both $t_c$
and $t_E$ share the same denominator $\ell_\hbar$, and for
physically realizable preparations $\delta_0 \leq a$, the
comparison $t_c \leq t_E$ reduces to $\delta_0 \leq a$ ---
the initial preparation uncertainty cannot exceed the characteristic
phase-space scale of the system. This is satisfied for any physically
realizable preparation. Therefore $t_c \leq t_E$: the inaccessibility
mechanism completes while classical mechanics remains exact. The
classical description is not assumed --- it is self-consistently
justified.
\subsection{Step 2: spaghettification}
\begin{figure}[t]
\centering
\includegraphics[width=\columnwidth,keepaspectratio]{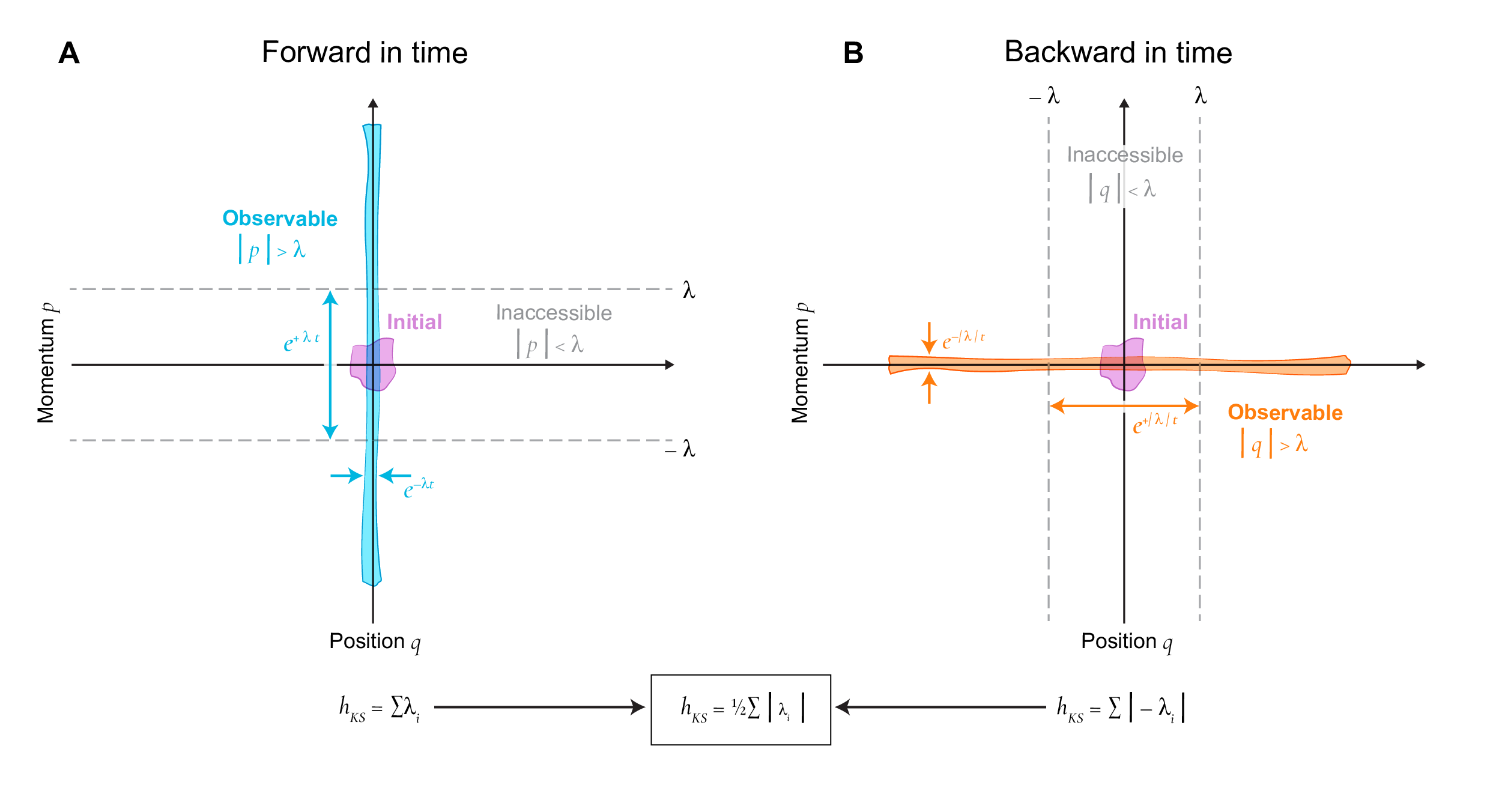}
\caption{\textbf{Spaghettification is time-symmetric.} \textbf{A:}
Forward evolution stretches phase space along unstable manifolds
while contracting along stable manifolds below quantum resolution
$\ell_\hbar$. \textbf{B:} Backward evolution reverses which manifold
expands. Both directions yield identical KS entropy production rate
$h_{KS} = \sum_{\lambda_i>0}\lambda_i = \frac{1}{2}\sum_i|\lambda_i|$.
The arrow of time is not in the dynamics.}
\label{fig:spaghetti}
\end{figure}

A Hamiltonian flow on phase space $\Gamma \in \mathbb{R}^{2N}$ is
symplectic: it preserves the symplectic form
$\omega = \sum_i dq_i \wedge dp_i$. The stability matrix
$\mathbf{M}(t)$ governing the separation of nearby trajectories
satisfies $\mathbf{M}^T J \mathbf{M} = J$, where $J$ is the
symplectic matrix. This forces its eigenvalues to pair as
$(\mu, 1/\mu)$, which implies the Lyapunov exponents pair as
$(\lambda, -\lambda)$~\cite{Oseledets1968}. This is the classical
reflection of the quantum Liouvillian pairing proved in Section~3:
symplecticity plays the same role as unitarity, enforcing the same
spectral symmetry through a different but analogous constraint.
The paired exponents mean phase space decomposes into an unstable
subspace (exponents $\lambda_i > 0$, trajectories diverge) and a
stable subspace (exponents $\lambda_i < 0$, trajectories converge).
Starting from an initial uncertainty $\delta_0$:
\begin{align}
\sigma_u(t) &\sim \delta_0\, e^{+\lambda t}
\qquad \text{(unstable: expands),} \\
\sigma_s(t) &\sim \delta_0\, e^{-\lambda t}
\qquad \text{(stable: contracts).}
\end{align}
Liouville's theorem guarantees that the total phase-space volume is
conserved: $\sigma_u \cdot \sigma_s \sim \delta_0^2$ at all times.
But the shape is radically altered. The initial uncertainty region
--- a sphere of radius $\delta_0$ --- is deformed into an
exponentially elongated, exponentially thin filament. This is
spaghettification: the same information, geometrically redistributed
(Fig.~\ref{fig:spaghetti}).
\subsection{Step 3: the critical time and operational inaccessibility}
Quantum mechanics sets a hard floor on distinguishability. Two states
whose separation in phase space falls below $\ell_\hbar \equiv
\sqrt{\hbar}$ in canonically scaled coordinates --- equivalently,
whose overlap in phase-space area falls below $\sim\hbar$ ---
are operationally indistinguishable: no physically admissible
measurement, preparation, or intervention can distinguish them
(Section~2).
Spaghettification drives the stable-direction width toward this
floor. The threshold is crossed at time $t_c$ defined by:
\begin{equation}
\sigma_s(t_c) = \delta_0\, e^{-\lambda t_c} = \ell_\hbar,
\end{equation}
giving:
\begin{equation}
\boxed{t_c = \frac{1}{\lambda}\ln\!\left(\frac{\delta_0}{\ell_\hbar}
\right).}
\end{equation}
Before $t_c$: the stable-direction separation exceeds $\ell_\hbar$.
The time-reversed microstate is distinct and in principle reachable
--- exponentially hard to reach, but reachable. After $t_c$: the
separation has fallen below $\ell_\hbar$. The time-reversed
microstate has become operationally indistinguishable from its
neighbors. No physical intervention can select it: the required
precision lies below $\ell_\hbar$, where quantum mechanics provides
no further resolution.
For typical molecular systems ($\delta_0/\ell_\hbar \sim 10^6$,
$\lambda \sim 10^{10}$\,s$^{-1}$), $t_c$ is on the order of
nanoseconds. The required reversal precision at $t = t_c$ is
$\sim e^{-30} \approx 10^{-13}$ of thermal momentum --- thirteen
orders of magnitude below thermal scales, and below any physically
realizable resolution. At macroscopic times ($t = 1$\,s) the
required precision is $\sim 10^{-10^{10}}$, which is not a number
that admits physical interpretation. The barrier is geometric.
The Loschmidt fidelity --- the probability that a state
time-reversed at $t$ returns to within $\delta_0$ of the original
--- undergoes a characteristic sigmoid decay around $t_c$
\cite{Pastawski2000,Jalabert2001,Sanchez2020}:
\begin{equation}
F(t) \approx \frac{1}{2}\,\mathrm{erfc}\!\left(
\frac{t - t_c}{\sqrt{2}\,\sigma_t}\right).
\end{equation}
The sigmoid shape --- a sharp threshold crossing, not exponential
leakage --- provides a characteristic signature consistent with
the geometric mechanism. It is observed in NMR Loschmidt echo
experiments and confirmed in the stadium-billiard simulation
of Section~6.
\subsection{Precise definition of accessibility}
\textbf{Definition (macroscopic accessibility).} A macrostate $M_2$
is \textit{accessible} from macrostate $M_1$ if there exists a
physically admissible intervention --- one acting only on
extensive, macroscopically resolvable degrees of freedom --- that
maps at least one microstate in $M_1$ to a microstate in $M_2$
under the system's Hamiltonian dynamics. A macrostate $M_2$
corresponding to entropy decrease is \textit{operationally
inaccessible} when the precision required to select the
appropriate microstate falls below $\ell_\hbar$.
This definition makes explicit what ``unreachable'' means: not
statistically suppressed, not computationally intractable, but
requiring a physical operation that quantum mechanics prohibits.
This definition formalizes physical reachability under
quantum-limited control, rather than redefining entropy itself.
Loschmidt's objection --- that time-reversed trajectories exist ---
is entirely correct. They are simply operationally inaccessible
in the precise sense defined here.
One difficulty still persists. Even if the time-reversed microstate
cannot be selected by any physically admissible operation, the
system might spontaneously evolve into it. Does this not constitute
a violation of the second law?
It does not. Under chaotic evolution with $\lambda > 0$, the
pre-image of any low-entropy macrostate develops exponentially
filamented structure with characteristic width below $\ell_\hbar$
after time $t_c$. No physically admissible operation --- bounded
by resolution $\ell_\hbar$ --- can prepare initial conditions from
this set. The corresponding microstates are outside the physically
reachable subset of phase space.
Moreover, coarse-graining and chaotic dynamics do not commute:
\begin{equation}
\pi \circ \phi_t \neq \phi_t^{\mathrm{macro}} \circ \pi,
\end{equation}
where $\phi_t$ is the Hamiltonian flow (the microscopic time
evolution map taking each microstate at time $0$ to its evolved
microstate at time $t$), and $\pi$ projects microstates to
macrostates. Once structure
crosses $\ell_\hbar$, the projection is many-to-one and the
macrostate partition is irreversibly deformed at the level of
physically distinguishable states. A trajectory that retraces its
microscopic evolution does not return to the same macrostate ---
the macrostate multiplicity structure has been altered.
In integrable systems ($\lambda = 0$), neither argument applies:
no filamentation, no deformation of $\pi$, no inaccessibility.
\subsection{Two equivalent pictures}
The mechanism can be described in two equivalent ways, in direct
analogy with the Schr\"{o}dinger and Heisenberg pictures of quantum
mechanics. Both are equally valid; the choice is one of language,
not physics.
\textbf{Picture 1 (state evolves, phase space fixed).} The
phase-space distribution --- equivalently the Wigner function ---
spaghettifies over a fixed phase space under chaotic evolution.
Stable directions contract until the distribution width falls below
$\ell_\hbar$. The time-reversed microstate remains a point in phase
space, but the distribution can no longer resolve it from its
neighbors. Inaccessibility is loss of resolution.
\textbf{Picture 2 (state fixed, phase space deforms).} The
operational metric on phase space --- the distance measure set by
distinguishability, consistent with the resolution scale $\ell_\hbar$
of Section~2 --- contracts along stable directions around a fixed
state. After time $t_c$, the metric distance between the
forward-evolved state and its time-reversed image falls below
$\ell_\hbar$. The two states are no longer metrically distinct.
Inaccessibility is metric collapse.
Both pictures preserve Liouville's theorem: symplectic volume ---
and hence the measure on phase space --- is unchanged. The arrow
of time lives in the metric, not the measure. In Picture~1 the
distribution becomes unresolvably thin; in Picture~2 the space
becomes unresolvably compressed. The physics is the same.
\subsection{What this argument does not claim}
This resolution of Loschmidt's paradox makes no appeal to:
probability or typicality (the time-reversed trajectories are not
rare, they are operationally inaccessible); observer-dependent
coarse-graining (the scale $\ell_\hbar$ is set by physics, not by
the observer's resolution); special initial conditions or
cosmological boundary conditions (the mechanism operates from any
initial state, in both time directions); or a breakdown of
microscopic reversibility (the dynamics remain exactly
time-symmetric, and the Lyapunov spectrum is $(\lambda,-\lambda)$
paired throughout). The asymmetry is geometric: stable manifolds
always contract, regardless of which direction is called forward.
\section{Quantitative Predictions}
The geometric mechanism makes predictions that are both quantitative
and testable. This section develops three: the irreversibility
timescale for a molecular gas, the scaling law for the critical time,
and the shape of the fidelity decay. Each is tied to a specific
experimental signature that distinguishes the geometric mechanism
from competing accounts of irreversibility.
\subsection{The irreversibility timescale}
For nitrogen gas at standard conditions, the maximum Lyapunov
exponent is estimated from the collision rate and mean free
path~\cite{Dorfman1999}:
\begin{equation}
\lambda \approx \frac{1}{\tau_{\rm coll}}
\ln\!\left(\frac{\lambda_{\rm mfp}}{d}\right)
\sim 10^{10}\,\text{s}^{-1},
\end{equation}
where $\tau_{\rm coll} \approx 1.8 \times 10^{-10}$\,s is the
mean collision time, $\lambda_{\rm mfp} \approx 8.6\times10^{-8}$\,m
is the mean free path, and $d = 3.7\times10^{-10}$\,m is the
molecular diameter. With $\delta_0/\ell_\hbar \sim 10^6$ for a
thermally prepared state, the critical time is:
\begin{equation}
t_c = \frac{1}{\lambda}\ln\!\left(\frac{\delta_0}{\ell_\hbar}\right)
\sim \frac{6\ln 10}{10^{10}} \sim 1\,\text{ns}.
\end{equation}
The required reversal precision at $t = t_c$ is:
\begin{equation}
\frac{\delta p_{\rm required}}{\delta p_{\rm thermal}}
\sim e^{-\lambda t_c} = \frac{\ell_\hbar}{\delta_0}
\sim 10^{-6},
\end{equation}
six orders of magnitude below the thermal momentum scale. At
$t = 1$\,s:
\begin{equation}
\frac{\delta p_{\rm required}}{\delta p_{\rm thermal}}
\sim e^{-10^{10}} \approx 10^{-10^{10}}.
\end{equation}
These are not statements about practical difficulty. They are
statements about phase-space geometry: the time-reversed microstate
lies below the quantum resolution floor, and no physically admissible
operation can resolve or select it.
\subsection{Scaling law for the critical time}
The geometric mechanism predicts a specific scaling:
\begin{equation}
t_c = \frac{1}{\lambda}\ln\!\left(\frac{\delta_0}{\ell_\hbar}\right).
\end{equation}
This has three testable features. First, $t_c$ scales as
$\lambda^{-1}$ --- systems with larger Lyapunov exponents become
irreversible faster. Second, $t_c$ grows logarithmically with the
ratio $\delta_0/\ell_\hbar$ --- doubling the preparation uncertainty
adds only $\lambda^{-1}\ln 2$ to the irreversibility time. Third,
$t_c$ is independent of ensemble size $M$ --- it is set by
trajectory geometry, not by statistics.
This last prediction is sharp. Accounts of irreversibility based
on ensemble averaging typically introduce an $M$-dependence into
the transition threshold. The geometric mechanism predicts strict
$M$-independence.
The stadium-billiard simulation is consistent with all three. The
critical time scales linearly with $\ln(\delta_0/\varepsilon)$ with
fitted slope $1/\lambda = 3.49 \pm 0.1$
(Fig.~\ref{fig:simulation}c). The $t_c$ contours are horizontal in
the $T$--$M$ plane across two orders of magnitude in $M$, with mean
variation below 2.5\% (Fig.~\ref{fig:simulation}b,e).
\subsection{Fidelity decay shape}
The geometric mechanism predicts that the Loschmidt fidelity
undergoes sigmoid decay around $t_c$~\cite{Jalabert2001,Pastawski2000}:
\begin{equation}
F(t) \approx \frac{1}{2}\,\mathrm{erfc}\!\left(
\frac{t - t_c}{\sqrt{2}\,\sigma_t}\right),
\end{equation}
where $\sigma_t$ reflects the spread of effective critical times
across Lyapunov directions and initial conditions. Assuming a
Gaussian distribution of effective critical times with variance
$\sigma_t^2$ across Lyapunov directions, the ensemble-averaged
fidelity takes this erfc form directly. The sigmoid shape
--- a sharp threshold crossing --- is the direct signature of
phase-space contraction below $\ell_\hbar$. It is qualitatively
distinct from purely exponential decay expected from noise-based or
perturbative accounts of irreversibility.
The stadium-billiard simulation shows sigmoid decay across all
tested values of $\delta_0$ (Fig.~\ref{fig:simulation}f). Three
decades of NMR Loschmidt echo experiments show perturbation-independent
decay~\cite{Sanchez2020,Pastawski2000}, consistent with a mechanism
driven by intrinsic phase-space geometry rather than external
imperfections.
\subsection{Quantum simulator test}
The clearest test is in quantum simulator platforms ---
trapped-ion or superconducting-qubit systems with $N = 50$--$100$
degrees of freedom and sufficient coherence time to observe $t_c$.
The prediction is unambiguous: fidelity decay should be sigmoid,
the critical time should scale as $\lambda^{-1}\ln(\delta_0/\ell_\hbar)$,
and the decay threshold should be independent of the number of
qubits used to represent the ensemble. Exponential decay, or
ensemble-size dependence of the threshold, would be inconsistent
with the geometric mechanism.
\section{Experimental Validation}
\begin{figure*}[t]
\centering
\includegraphics[width=\textwidth,keepaspectratio]{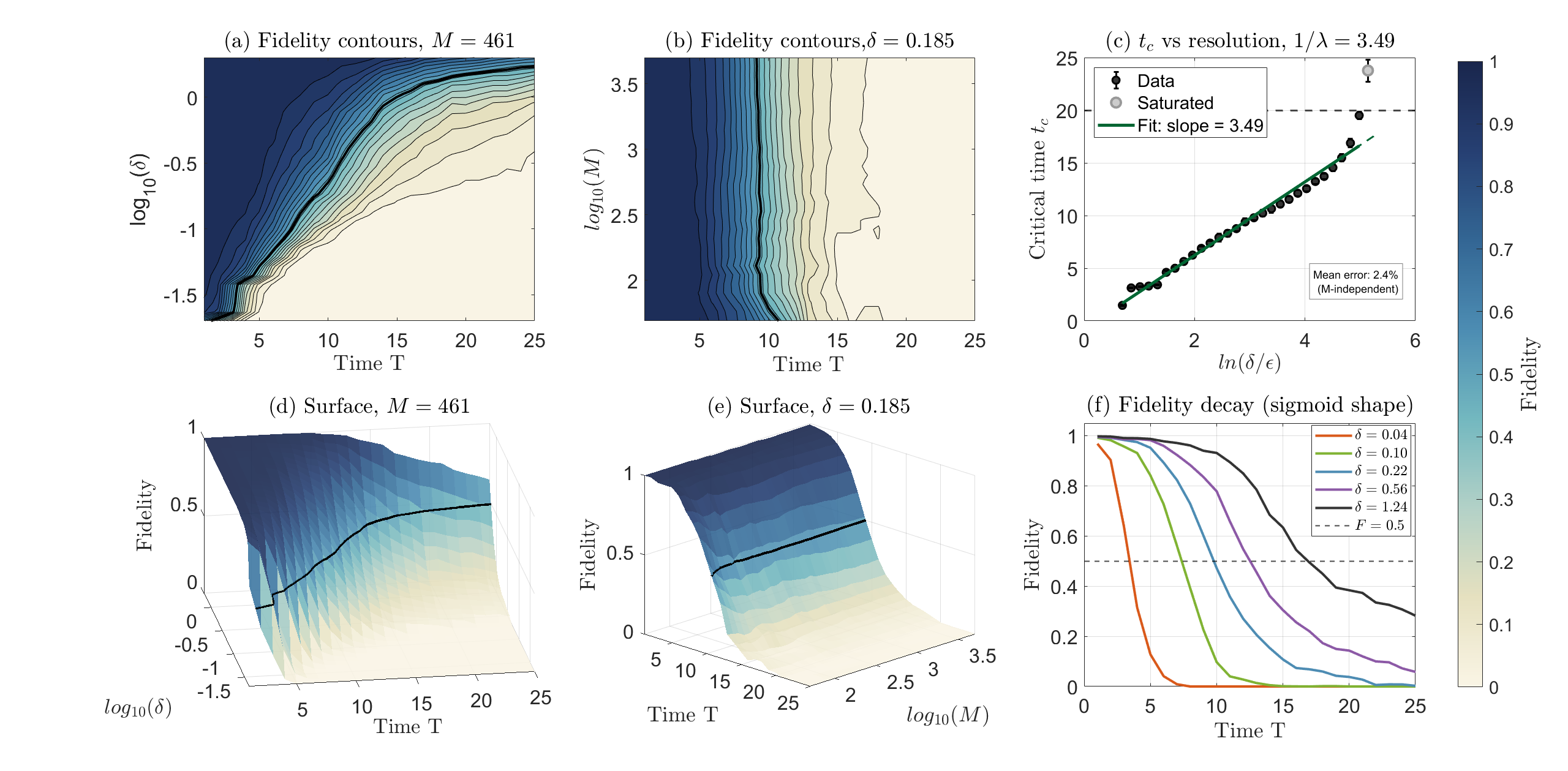}
\caption{\textbf{Stadium-billiard Loschmidt echo simulation.}
\textbf{(a)} Fidelity contours vs time $T$ and coarse-graining
$\delta$ at fixed $M=461$. The diagonal ridge marks the
reversible-to-irreversible transition.
\textbf{(b)} Fidelity contours vs $T$ and $M$ at fixed
$\delta=0.185$; horizontal contours confirm $M$-independence.
\textbf{(c)} Critical time $t_c$ vs $\ln(\delta/\varepsilon)$:
linear scaling with slope $1/\lambda = 3.49$. Gray points exceed
simulation window (saturation artifact).
\textbf{(d,e)} Three-dimensional surfaces for panels (a,b).
\textbf{(f)} Fidelity decay curves for different $\delta$ showing
characteristic sigmoid shape. Perturbation $\varepsilon = 0.01$.}
\label{fig:simulation}
\end{figure*}
Section~5 identified three testable signatures of the geometric
mechanism: sigmoid fidelity decay, logarithmic scaling of $t_c$,
and independence of $t_c$ from ensemble size. This section shows
that all three are consistent with existing experiments and a
dedicated numerical simulation.
\subsection{Loschmidt echo experiments}
The Loschmidt echo protocol provides a direct operational test of
reversibility~\cite{PastawskiScholarpedia}. An initial state is
prepared, evolved forward under Hamiltonian $H$ for time $T$, then
evolved backward under $-H$ (implemented via NMR pulse sequences in
many-body spin systems). The echo amplitude $M(T)$ measures the
return fidelity --- how faithfully the system retraces its path.
If irreversibility were driven by external imperfections --- pulse
errors, environmental noise, finite perturbations to the Hamiltonian
--- the echo decay rate would scale with the perturbation strength.
This is not what is observed. Pastawski and collaborators established
over three decades of experiments that the decay persists at a rate
set by intrinsic Hamiltonian properties, independent of perturbation
strength~\cite{Pastawski2000,Jalabert2001,Sanchez2020,Levstein1998}.
This perturbation-independent regime is precisely what the geometric
mechanism predicts: $t_c$ depends on $\lambda$ and $\delta_0/\ell_\hbar$,
neither of which is set by the perturbation.
The observed decay is consistent with a threshold-crossing sigmoid
rather than exponential leakage~\cite{Jalabert2001,Sanchez2020}.
This provides a characteristic signature of phase-space contraction
below $\ell_\hbar$: a sharp transition when the stable manifolds
cross the quantum resolution floor, not a continuous degradation
from noise.
\subsection{Stadium-billiard simulation}
To test the geometric mechanism in a controlled setting with known
Lyapunov exponent, we simulated the Loschmidt echo in the Bunimovich
stadium billiard~\cite{Bunimovich1979} --- a paradigmatic chaotic
system whose statistical properties are well characterised. An
ensemble of $M$ particles evolves forward for time $T$ under exact
Hamiltonian dynamics, then undergoes velocity reversal with a small
perturbation $\varepsilon = 0.01$ to the boundary radius, then
evolves backward for time $T$. The fidelity $F(T,\delta)$ measures
the fraction of particles returning within coarse-graining scale
$\delta$ of their initial phase-space position. Full simulation
details are provided in the Supplemental Material~\cite{SupMat}.
The results are consistent with all three predictions of Section~5.
\textit{Sigmoid decay.} Fidelity decays as a sigmoid rather than
exponentially across all tested values of $\delta$
(Fig.~\ref{fig:simulation}f). This is consistent with a
threshold-crossing picture: the transition from reversible to
irreversible behavior is sharp, not a gradual leakage. A purely
exponential decay would be difficult to reconcile with the geometric
mechanism.
\textit{Logarithmic scaling of $t_c$.} The critical time scales
linearly with $\ln(\delta/\varepsilon)$, with fitted slope
$1/\lambda = 3.49 \pm 0.1$ (Fig.~\ref{fig:simulation}c), consistent
with $t_c = \lambda^{-1}\ln(\delta_0/\ell_\hbar)$. Here the
perturbation scale $\varepsilon$ plays the role of an effective
resolution scale, analogous to $\ell_\hbar$ in the quantum case.
Error bars (mean variation 2.4\%) confirm that $t_c$ is a geometric
property of the trajectories, not a statistical artifact.
\textit{$M$-independence.} The $t_c$ contours are horizontal in
the $T$--$M$ plane across two orders of magnitude in $M$, with
mean variation below 2.5\% (Fig.~\ref{fig:simulation}b,e). The
transition time is determined by the coarse-graining scale $\delta$
alone, not by how many particles are used to represent the ensemble.
The stadium billiard isolates the simplest possible case: a single
particle, chaos from boundary geometry alone, no inter-particle
interactions. The mechanism is visible in its cleanest form here.
Many-body systems with particle--particle collisions have larger
Lyapunov exponents and reach $t_c$ faster, which strengthens rather
than weakens the argument.
\section{Relation to Prior Approaches}
\textbf{Molecular chaos (Boltzmann~\cite{Boltzmann1872}).}
The Stosszahlansatz assumes pre-collision velocity decorrelation,
introducing time asymmetry at the level of initial conditions.
It does not identify a dynamical mechanism that enforces this
asymmetry. The present work provides a concrete mechanism for this:
chaotic contraction drives correlations below $\ell_\hbar$, making
them operationally inaccessible rather than merely assumed absent.
\textbf{Coarse-graining (Gibbs, Penrose~\cite{Penrose1970}).}
Entropy is defined relative to a finite partition of phase space.
The open question is what fixes the physically relevant cell size.
The present work identifies a physically motivated answer: the
minimum cell is set by $\ell_\hbar$, not by observer convention.
\textbf{Information-theoretic entropy (Jaynes~\cite{Jaynes1957a}).}
Entropy measures missing information. This is consistent with the
present framework but does not explain why information loss is
directional under time-symmetric dynamics. The geometric mechanism
provides the missing dynamical explanation.
\textbf{Past Hypothesis (Albert, Carroll~\cite{Albert2000,Carroll2010,Zeh2007}).}
Cosmological boundary conditions explain the low-entropy early
universe. The present work addresses a different question: given
any initial state prepared at finite resolution, when does dynamical
evolution render time reversal physically inaccessible? The answer
is independent of cosmological history.
\textbf{Typicality (Lebowitz~\cite{Lebowitz1993,Goldstein2001}).}
Entropy increase is typical: the overwhelming majority of microstates
consistent with a low-entropy macrostate evolve toward higher entropy.
This is correct. But typicality is a measure-theoretic statement ---
it does not explain why entropy-decreasing trajectories are
dynamically unreachable from physically preparable states. The
present work makes this step operational.
\textbf{Fluctuation theorems
(Gallavotti--Cohen~\cite{Gallavotti1995,Evans1993}).}
The fluctuation theorem quantifies the probability ratio of
entropy-increasing to entropy-decreasing trajectories in driven
systems. This is compatible with the present framework: where
the fluctuation theorem describes probabilities, the present work
identifies the geometric condition under which entropy-decreasing
trajectories become operationally inaccessible.
\textbf{Lanford's derivation~\cite{Lanford1975}.}
Lanford rigorously derived the Boltzmann equation from Hamiltonian
mechanics for $t < \frac{1}{5}\tau_{\rm coll}$, but assumed
molecular chaos as an initial condition. The present work
identifies a mechanism by which this assumption becomes dynamically
enforced at $t > t_c$: stable manifold contraction below $\ell_\hbar$
renders pre-collision correlations inaccessible.
\textbf{Quantum kicked rotator
(Shepelyansky~\cite{Shepelyansky1983}).}
Dynamical localization in the periodically driven kicked rotator
renders the quantum evolution effectively reversible at long times,
and has been raised as a counterexample to geometric inaccessibility.
It does not apply here for two reasons. First, the kicked rotator
is non-autonomous: energy is continuously injected by the periodic
drive. The present framework applies to autonomous,
energy-conserving systems. Second, dynamical localization operates
after $t_E$, in the regime where quantum interference dominates.
The geometric mechanism operates entirely by $t_c \leq t_E$, before
localization becomes relevant. This separation is quantitative: for
the kicked rotator with kicking strength $K$ and effective Planck
constant $\hbar_{\rm eff}$, the Ehrenfest time scales as
$t_E \sim \lambda^{-1}\ln(1/\hbar_{\rm eff})$ while the
localization time scales as
$t_{\rm loc} \sim K^2/\hbar_{\rm eff}$~\cite{Shepelyansky1983}.
In the semiclassical regime $\hbar_{\rm eff} \to 0$,
$t_{\rm loc}/t_E \sim K^2/(\hbar_{\rm eff}\ln(1/\hbar_{\rm eff}))
\to \infty$: localization sets in parametrically later than $t_E$,
leaving the geometric mechanism intact.
\section{Conclusion}
This paper makes two contributions.
First, for any Hamiltonian satisfying $H(t) = H(-t)$,
the eigenvalues of the quantum Liouvillian $\mathcal{L}_t =
-\frac{i}{\hbar}[H(t),\cdot]$ come in pairs $(\lambda,-\lambda)$
at every $t$. Via a rigorous Krylov complexity construction ---
proving Hermiticity of $[H,\cdot]$ under the Hilbert--Schmidt
inner product, the three-term Lanczos recurrence, and the
invariance of the off-diagonal coefficients $b_n$ under
$\hat{\mathcal{L}} \to -\hat{\mathcal{L}}$ (Favard's theorem)
--- we establish:
\begin{equation}
\lambda_L^{\rm forward} = \lambda_L^{\rm backward}.
\end{equation}
The rate at which initially distinguishable states become
operationally indistinguishable is identical in both time directions.
The arrow of time is not in the dynamics.
Secondly, chaotic Hamiltonian evolution contracts phase-space
structure along stable directions until the separation between
initially distinguishable states falls below the quantum resolution
scale $\ell_\hbar$. This occurs at:
\begin{equation}
t_c = \frac{1}{\lambda}\ln\!\left(\frac{\delta_0}{\ell_\hbar}\right),
\end{equation}
which satisfies $t_c \leq t_E$ for any physically realizable
preparation --- the mechanism operates within the semiclassical
regime where classical phase space is the valid description.
After $t_c$, the time-reversed microstate becomes operationally
inaccessible: the required precision lies below $\ell_\hbar$,
where quantum mechanics provides no further resolution.
The argument then flows: entropy measures our uncertainty over which
distinguishable quantum state the system occupies; the minimum
distinguishable cell is set by $\ell_\hbar$, not by observer
convention (Section~2). The quantum Liouvillian proof shows that
the rate of distinguishability loss is time-symmetric --- the
dynamics contribute no arrow (Section~3). The geometric mechanism
then identifies where the arrow does come from: chaotic contraction
drives stable-direction separations below $\ell_\hbar$ at time
$t_c$, rendering the time-reversed microstate unreachable
(Section~4). Quantitative estimates give $t_c \sim 1$\,ns for a
molecular gas, with required reversal precision falling to
$\sim 10^{-10^{10}}$ of thermal momentum at macroscopic times
(Section~5). Three experimental signatures --- sigmoid fidelity
decay, logarithmic scaling of $t_c$, and $M$-independence ---
are consistent with Loschmidt echo experiments and confirmed in a
stadium-billiard simulation (Section~6).
The argument works because it separates two questions that are
usually conflated. The first is dynamical: does the Hamiltonian
carry a preferred time direction? It does not --- the Liouvillian
proof establishes this at the quantum level, and the symplectic
structure of classical mechanics reflects the same fact. The second
is geometric: are time-reversed microstates physically reachable?
After $t_c$ they are not --- the geometric mechanism identifies the
precise condition under which they fall below the quantum resolution
floor.
Loschmidt's objection was correct: time-reversed trajectories
exist, Hamilton's equations are time-symmetric, and Liouville's
theorem preserves phase-space measure. Nothing here contradicts
any of that. The result does not remove time-reversed trajectories
from the theory; it identifies the conditions under which they
cease to be physically accessible. The resolution is that existence
and accessibility are different things. After $t_c$, the
time-reversed microstate is real but unreachable. The second law
does not require a breakdown of microscopic reversibility. It
requires only that chaotic dynamics, given any finite preparation
uncertainty, will drive stable manifolds below the quantum
resolution scale --- and this, for any macroscopic system, follows
generically. In this sense, irreversibility emerges as a constraint
on physical accessibility rather than a modification of the
underlying dynamics.
\subsection{Poincar\'{e} recurrence}
Poincar\'{e} recurrences are formally allowed and occur on
timescales of order $e^{S/k_B}$, far beyond any physically
relevant window. The geometric mechanism does not forbid
recurrences; it operates on the timescales where thermodynamics
applies. Even if a system were to pass near its initial macrostate
after a Poincar\'{e} time, no macroscopic operation could verify
or exploit the microscopic trajectory reversal required to violate
the second law. Recurrence and geometric inaccessibility are
compatible: one is a statement about formal measure, the other
about physical accessibility.

\end{document}


\title{Supplemental Material: Quantum Inaccessibility}

\author{Ira Wolfson\\
\small Department of Electrical and Electronic Engineering,\\
\small Braude Academic College of Engineering, Karmiel 2161002, Israel\\
\small \texttt{wolfsoni@braude.ac.il}}

\date{}
\maketitle

\section{Canonical Nondimensionalization and the Definition of
$\ell_\hbar$}

The main text uses $\ell_\hbar \equiv \sqrt{\hbar}$ as the quantum
resolution scale, but the precise meaning requires care: $\hbar$ has
units of action (J$\cdot$s), so $\sqrt{\hbar}$ is not a length.
This section defines $\ell_\hbar$ precisely in canonically
nondimensionalized coordinates.

Choose a reference action scale $\hbar$ and define dimensionless
canonical coordinates:
\begin{equation}
\tilde{q} = \frac{q}{q_*}, \qquad \tilde{p} = \frac{p}{p_*},
\qquad q_* p_* = \hbar,
\end{equation}
where $q_*$ and $p_*$ are dimensional scales (e.g.\ thermal de
Broglie wavelength and thermal momentum). In these coordinates the
Heisenberg uncertainty principle reads:
\begin{equation}
\Delta\tilde{q}\cdot\Delta\tilde{p} \geq \frac{1}{2}.
\end{equation}
The minimum physically distinguishable phase-space cell has area
$\sim 1/2$ in $(\tilde{q},\tilde{p})$ coordinates. We define:
\begin{equation}
\ell_\hbar \equiv \mathcal{O}(1) \quad \text{in dimensionless
canonical coordinates,}
\end{equation}
so that $\Delta\tilde{q} \sim \Delta\tilde{p} \sim \ell_\hbar$ for
a minimum-uncertainty state. The threshold condition
$\sigma_s(t_c) = \ell_\hbar$ is then a statement in canonical
invariants: the stable-direction width (measured in units of the
quantum cell scale) reaches the quantum floor. The choice of $q_*$
and $p_*$ drops out of the critical time formula, provided
$q_*p_* = \hbar$.

Any preparation made by macroscopic means satisfies
$\delta_0 \gg \ell_\hbar$ in these coordinates, since a macroscopic
apparatus cannot prepare states at the quantum resolution limit.

\section{Stability Matrix and Lyapunov Pairing:
Rigorous Derivation}

This section provides the complete derivation supporting the
classical Lyapunov pairing claim in Section~4 of the main text.

\subsection{Setup}

Consider a Hamiltonian system with $N$ degrees of freedom,
phase-space coordinates $\Gamma \in \mathbb{R}^{2N}$, and flow
$\phi_t: \Gamma_0 \mapsto \Gamma(t)$ generated by Hamilton's
equations:
\begin{equation}
\dot{\Gamma} = J\nabla_\Gamma H, \qquad
J = \begin{pmatrix} 0 & I \\ -I & 0 \end{pmatrix}.
\end{equation}
For a reference trajectory $\Gamma^*(t)$, nearby deviations
$\delta\Gamma(t)$ evolve as:
\begin{equation}
\frac{d}{dt}\delta\Gamma = \mathbf{L}(t)\,\delta\Gamma,
\qquad
\mathbf{L}(t) = J\,H''(\Gamma^*(t)),
\end{equation}
where $H''$ is the Hessian of $H$ evaluated along $\Gamma^*(t)$.
The solution is $\delta\Gamma(t) = \mathbf{M}(t)\,\delta\Gamma(0)$,
where the stability matrix satisfies:
\begin{equation}
\dot{\mathbf{M}} = \mathbf{L}(t)\mathbf{M}, \qquad \mathbf{M}(0) = I.
\end{equation}

\subsection{Symplecticity of $\mathbf{M}(t)$}

\textbf{Proposition.} $\mathbf{M}(t)^T J\,\mathbf{M}(t) = J$
for all $t$.

\textit{Proof.} Define $\mathbf{A}(t) = \mathbf{M}^T J\mathbf{M}$.
At $t=0$, $\mathbf{A}(0) = J$. Differentiating:
\begin{equation}
\dot{\mathbf{A}} = \dot{\mathbf{M}}^T J\mathbf{M}
+ \mathbf{M}^T J\dot{\mathbf{M}}
= \mathbf{M}^T(\mathbf{L}^T J + J\mathbf{L})\mathbf{M}.
\end{equation}
Using $\mathbf{L} = JH''$, $J^T = -J$, and symmetry of $H''$:
\begin{equation}
\mathbf{L}^T = (JH'')^T = H''^T J^T = H''(-J) = -H''J,
\end{equation}
so:
\begin{equation}
\mathbf{L}^T J = -H''J^2 = H'', \qquad
J\mathbf{L} = J^2 H'' = -H''.
\end{equation}
Therefore:
\begin{equation}
\mathbf{L}^T J + J\mathbf{L} = H'' - H'' = 0,
\end{equation}
so $\dot{\mathbf{A}} = 0$. Since $\mathbf{A}(0) = J$, we have
$\mathbf{M}^T J\mathbf{M} = J$ for all $t$. $\blacksquare$

\subsection{Eigenvalue pairing $(\mu, 1/\mu)$}

\textbf{Corollary.} If $\mu$ is an eigenvalue of $\mathbf{M}(t)$,
then so is $1/\mu$.

\textit{Proof.} Let $\mathbf{M}v = \mu v$. Then immediately
$\mathbf{M}^{-1}v = \mu^{-1}v$, so $\mu^{-1}$ is an eigenvalue
of $\mathbf{M}^{-1}$. From the symplectic condition and
$J^{-1} = -J$:
\begin{equation}
\mathbf{M}^{-1} = J^{-1}\mathbf{M}^T J = -J\mathbf{M}^T J.
\end{equation}
This shows $\mathbf{M}^{-1}$ is similar to $\mathbf{M}^T$
(conjugation by $-J$). Similar matrices share eigenvalues, and
$\mathbf{M}^T$ shares eigenvalues with $\mathbf{M}$ (transposition
does not change the characteristic polynomial). Therefore
eigenvalues of $\mathbf{M}^{-1}$ = eigenvalues of $\mathbf{M}$.
Since $\mu^{-1}$ is an eigenvalue of $\mathbf{M}^{-1}$, it is
also an eigenvalue of $\mathbf{M}$. $\blacksquare$

\subsection{Lyapunov pairing $(\lambda, -\lambda)$}

The Lyapunov exponents are the asymptotic growth rates of the
singular values of $\mathbf{M}(t)$:
\begin{equation}
\lambda_i = \lim_{t\to\infty}\frac{1}{t}\ln\sigma_i(t).
\end{equation}
We show directly that the singular values pair as $(\sigma, 1/\sigma)$.
From the symplectic condition, $\mathbf{M}^{-1} = J^{-1}\mathbf{M}^T J$.
Since $J^T J = I$ (that is, $J$ is orthogonal), pre- and
post-multiplication by $J^{\pm 1}$ preserves singular values.
Therefore the singular values of $\mathbf{M}^{-1}$ equal those
of $\mathbf{M}^T$, which equal those of $\mathbf{M}$. But the
singular values of $\mathbf{M}^{-1}$ are $\{1/\sigma_i(\mathbf{M})\}$.
Hence $\{1/\sigma_i\} = \{\sigma_i\}$: the singular values pair as
$(\sigma, 1/\sigma)$, and therefore the Lyapunov exponents pair as
$(\lambda, -\lambda)$. The Oseledets
multiplicative ergodic theorem~\cite{Oseledets1968} guarantees
these limits exist for almost all trajectories in ergodic systems.

The Kolmogorov-Sinai entropy is therefore time-symmetric at the
classical level:
\begin{equation}
h_{KS}^{\rm forward} = \sum_{\lambda_i > 0}\lambda_i
= \sum_{\lambda_i > 0}|\lambda_i|
= h_{KS}^{\rm backward}.
\end{equation}
This is the classical reflection of the quantum Liouvillian result
proved in Section~3 of the main text.

\section{Numerical Simulation: Stadium Billiard}

\subsection{System}

The Bunimovich stadium billiard~\cite{Bunimovich1979} consists of
two semicircular caps of radius $R$ connected by straight walls of
length $2a$. We use $R = a = 1$ (natural units), particle speed
$|v| = 1$. The stadium is a paradigmatic chaotic system with
well-characterised Lyapunov exponent $\lambda \approx 0.29$
(fitted from simulation, consistent with literature values).

\subsection{Protocol}

\begin{enumerate}
\item Initialize $M$ particles with positions uniformly distributed
inside the stadium and random velocity directions, $|v| = 1$.
\item Evolve forward for time $T$ under exact collision dynamics
with boundary radius $R$.
\item Reverse all velocities: $\mathbf{v} \to -\mathbf{v}$.
\item Evolve under perturbed dynamics with boundary radius
$R' = R(1 + \varepsilon)$, $\varepsilon = 0.01$.
\item Measure fidelity:
\begin{equation}
F(T,\delta) = \frac{1}{M}\sum_{i=1}^{M}
\Theta\!\left(\delta - d_i(T)\right),
\end{equation}
where $d_i(T) = \sqrt{|\mathbf{r}_i^{\rm final}
- \mathbf{r}_i^{\rm initial}|^2
+ |\mathbf{v}_i^{\rm final} - \mathbf{v}_i^{\rm initial}|^2}$
is the Euclidean phase-space distance and $\Theta$ is the
Heaviside function.
\end{enumerate}

The perturbation $\varepsilon$ plays the role of an effective
resolution scale, analogous to $\ell_\hbar$ in the quantum case:
$t_c = \lambda^{-1}\ln(\delta/\varepsilon)$.

\subsection{Algorithm}

An event-driven algorithm with exact collision detection avoids
numerical integration errors. A collision table stores the
predicted time to each wall for each particle. At each step, the
earliest collision is found, all particles are advanced to that
time, the colliding particle's velocity is reflected off the wall
normal, and the table is updated for that particle only. Flat
walls use linear intersection; circular caps use the positive root
of the quadratic $|\mathbf{r} + \mathbf{v}t - \mathbf{c}|^2 = R^2$.

\subsection{Parameters}

\begin{center}
\begin{tabular}{ll}
\toprule
Parameter & Value \\
\midrule
Stadium geometry & $R = 1$, $a = 1$ \\
Particle speed & $|v| = 1$ \\
Perturbation & $\varepsilon = 0.01$ \\
Time range & $T \in [1, 25]$ (25 values) \\
Ensemble sizes & $M \in [49, 4999]$ (30 log-spaced) \\
Coarse-graining & $\delta \in [0.02, 2.0]$ (30 log-spaced) \\
Random seed & 42 \\
\bottomrule
\end{tabular}
\end{center}

\subsection{Results}

\textbf{Sigmoid decay.} Fidelity curves (Main Text Fig.~2f)
exhibit sigmoid shape across all tested values of $\delta$,
consistent with threshold crossing rather than exponential decay.

\textbf{Critical time scaling.} $t_c$ (defined as $F = 0.5$)
scales linearly with $\ln(\delta/\varepsilon)$ with fitted slope
$1/\lambda = 3.49 \pm 0.1$, giving $\lambda \approx 0.29$.
Mean variation 2.4\% confirms $t_c$ is a geometric property of
trajectories.

\textbf{$M$-independence.} Mean relative variation of $t_c$
across $M \in [49, 4999]$ is below 2.5\%. The $t_c$ contours
are horizontal in the $T$--$M$ plane (Main Text Fig.~2b,e).

\subsection{Saturation artifact}

At large $\delta$, the predicted $t_c$ exceeds the simulation
window $T_{\rm max} = 25$. These cases (gray in Main Text
Fig.~2c) are excluded from the linear fit. The cutoff
$t_c > 0.8\,T_{\rm max} = 20$ separates the physical scaling
regime from the finite-time artifact.

\subsection{Code availability}

Python simulation code is available at\\
\url{https://github.com/beastraban/Stadium-Billiard-Loschmidt-Echo}.